\documentclass[12pt,a4paper]{article}
\usepackage{amsmath,amsfonts,amssymb}
\usepackage[english]{babel}
\let\p\partial

\newcommand{\op}[2]{\vskip 1mm{\bf #1.}{\ #2}\vskip 1mm}

\begin{document}
\title{Integrals of open 2D lattices} %
\author{Dmitry K. Demskoi \\
Academia Sinica, Taipei, Taiwan\\
{\small e-mail: demskoi@math.sinica.edu.tw }
}
\date{}
\maketitle
\begin{abstract}
We present an explicit formula for integrals of the open 2D Toda lattice of type $A_n$. This formula is applicable for various reductions of this lattice. To illustrate the concept we find integrals of the Toda $G_2$ lattice.
We also reveal a connection between the open Toda $A_n$ and Shabat-Yamilov lattices.
\end{abstract}
\section{Introduction}
The most well known representatives of the class of exactly solvable hyperbolic systems are open 2D Toda lattices
\begin{equation}
	u_{i,tx}=\exp\left(A^i_j u^j\right), \ \ j=1,\dots,n,
	\label{alltodas}
\end{equation}
where $(A_j^i)$ is the Cartan matrix of a simple Lie algebra. A general method of integration of such systems was proposed by Leznov and Saveliev \cite{leznov}. The version of lattice (\ref{alltodas}) corresponding to classical series $A_n$ was known to Darboux who also found its general solution.
Exactly solvable systems have a few characteristic properties that set them apart from the multitude of all other systems. These include in particular: finiteness of chains of generalized Laplace invariants \cite{zhibsok,zhib}, presence of non-trivial integrals, and generalized symmetries. It is also known that the latter two structures are related to each other by means of a differential operator mapping integrals to symmetries \cite{sokstarts}. 
Systems (\ref{alltodas}) have long been known to possess the complete sets of integrals \cite{shabyam}, however, the explicit formulas for them have never been presented apart from a few particular cases. In this paper we suggest a solution of this problem for the $A_n$ Toda lattice and its reductions.

The simplest hyperbolic integrable equation is the d'Alembert equation
\begin{equation}
	w_{tx}=0.
	\label{wave}
\end{equation}
It is not obvious, however, how this equation can be generalized to the case of higher order equations or systems of equations when it is written in this form.
If we introduce the new dependent variable $w=\log(u)$, then (\ref{wave}) becomes
\begin{equation}
	\mbox{det}\left(\begin{array}{cc}
	u & u_t \\
	u_x & u_{tx}
	\end{array}\right)=0.
	\label{wave2}
\end{equation}
The obvious generalisation of (\ref{wave2}) is the $2n$-th order equation
\begin{equation}
	\mbox{det}\left(\begin{array}{cccc}
	u & u_t&\dots & u_{t\dots t} \\
	u_x & u_{tx} & \dots &  u_{t\dots tx} \\
	\vdots & \vdots &\ddots &\vdots \\
	u_{x\dots x} & u_{tx\dots x} & \dots & u_{t\dots tx\dots x}
	\end{array}\right)=0
	\label{2nth}
\end{equation}
where $u_{t\dots tx\dots x}=\p_x^n \p_t^n u$. For brevity we write equation (\ref{2nth}) as
$$
W_{n+1}(u)=0.
$$
This equation is central for our further considerations.
In the sequel it will be referred as the higher d'Alembert equation. Equation (\ref{2nth}) can be viewed as the zero condition for the Wronskians: 
$$W(u,u_x,\dots,u_{x\dots x})(t)=0,\ \ W(u,u_t,\dots,u_{t\dots t})(x)=0.$$
From this we can deduce the general solution
\begin{equation}
	u=X_1(x)T_1(t)+X_2(x)T_2(t)+\dots +X_n(x)T_n(t),
	\label{scalgensol}
\end{equation}
where $X_i, T_i$ are arbitrary functions. 

Apparently there are different ways of writing (\ref{2nth}) as a system of equations. It was shown by Darboux that quantities $W_j(u)$ satisfy the recurrent relation
\begin{equation}
	(\ln W_j)_{xt}=W_{j-1}W_{j+1}W_j^{-2},\ \ W_0=1,\ \ W_1=u.
	\label{rekur}
\end{equation}
This relation initially appeared in connection with studying the Laplace invariants of hyperbolic equations.
Upon introducing the new quantities
\begin{equation}
		W_j(u)=\exp(w_j),
		\label{transf}
\end{equation}
equation (\ref{2nth}) is transformed into the system
\begin{equation}
w_{j,xt}=\exp(w_{j-1}-2w_{j}+w_{j+1}),\ \ w_{n,xt}=0,\ \ j=1\dots n-1
	\label{todaa1}
\end{equation}
with the boundary condition $w_0=0$.
Note that we can eliminate $w_n$ from this system by means of the transformation
$$ w_{n-1}\to w_{n-1}+\tfrac{2}{3}w_n,\ \ w_{n-2}\to w_{n-2}+\tfrac{1}{3}w_n.$$
The resulting system reads
\begin{equation}
w_{j,xt}=\exp(w_{j-1}-2w_{j}+w_{j+1}),\ \  w_0=w_n=0, \, j=1\dots n-1.
	\label{_todaa1}
\end{equation}
System (\ref{_todaa1}) is often referred as the open (finite, non-periodic) $A_{n-1}$ Toda lattice. Therefore equation (\ref{2nth}) and the open $A_n$ Toda lattice are arguably the simplest generalisations of the d'Alembert equation for higher order equations and systems of equations. We may wonder if there are other systems related to equation (\ref{2nth}) which have reasonably compact form? Apparently any other system reducible to equation (\ref{2nth}) must also be related to (\ref{todaa1}). The other question we are interested in is: How other structures related the solvability of the higher d'Alembert equation are related to those of corresponding systems of equations. Obviously the integrals of equation (\ref{2nth}) are integrals of (\ref{todaa1}) as well. Surprisingly the formula for the scalar equation is much simpler than for the corresponding system of equations.

\section{Integrals of open Toda Lattices}
Let us review some properties of higher d'Alembert equation.  
We have already indicated its general solution due to Darboux, now we want to show that it also possesses $n+n$ independent integrals.
Note that due to the symmetry $x\leftrightarrow t$ it suffices to present $t-$integrals only. Recall that a function of unknowns and their derivatives is called $t-$integral if it satisfies the characteristic equation $D_t \omega=0$, see \cite{zhibsok} for detailed exposition.
\op{Proposition}{Equation (\ref{2nth}) admits $n$ independent $t-$integrals of the form
\begin{equation}
	\omega_i=\frac{W_{n,i}}{W_{n}}, 
	\label{scalints}
\end{equation}
where $W_{n,i}$ is the determinant derived from $W_{n}$ by replacing its $i$-th row by 
$$(\p_x^n u, \p_x^n\p_t u, \p_x^n\p_t^2 u, \dots, \p_x^n\p_t^{n-1} u ).$$
}
Perhaps the easiest way to prove this statement is to show that integrals (\ref{scalints}) become functions of one variable after substituting expression of general solution (\ref{scalgensol}) into (\ref{scalints}). Indeed, after substituting we get
$$
\begin{array}{c}
W_{n,i}=W(T_1,\dots,T_n)(t)\sum_j X_j^{(n)}C_{ij}(x),\\[1mm] 
W_n=W(X_1,\dots,X_n)(x)\,W(T_1,\dots,T_n)(t),
\end{array}
$$
where $C_{ij}(x)$ is the cofactor of the entry $(W(X_1,\dots,X_n)(x))_{ij}$. Therefore the integrals of (\ref{2nth}) are parametrized by functions $X_i(x)$ the following way
$$
\omega_i=\frac{\sum_j X_j^{(n)}C_{ij}(x)}{W(X_1,\dots,X_n)(x)}.
$$
Independence of these integrals follows from formula (\ref{scalints}) itself.
\op{Remark}{Formula (\ref{scalints}) provides us with the explicit expression for integrals not only of (\ref{2nth}), but of (\ref{todaa1}) as well. Note that $u=W_1=\exp(w_1)$, and hence $\omega_i$ can be expressed in terms of the single quantity $w_1=\log(u)$. }

{ Instead of (\ref{2nth}) and (\ref{todaa1}) we could have started with system (\ref{_todaa1}). One can show that the latter is equivalent to the scalar equation \cite{leznov}
$$
W_{n}(u)=(-1)^{n(n-1)/2}.
$$
In this case the formulas for integrals should be modified:
\begin{equation}
	\omega_i=\frac{W^*_{n,i}}{W_{n}}, 
	\label{scalintsm}
\end{equation}
where $W^*_{n,i}$ is the determinant derived from $W_{n}$ by replacing its $i$-th row by 
$$(\p_x^{n+1} u, \p_x^{n+1}\p_t u, \p_x^{n+1}\p_t^2 u, \dots, \p_x^{n+1}\p_t^{n-1} u ).$$
}
Formulas (\ref{scalints}) and (\ref{scalintsm}) can be used for finding integrals of various lattices derived from equation (\ref{2nth}). This includes
the 2D Toda lattice corresponding to Cartan matrix of the Lie Algebra $A_n$ and its reductions, other version of the 2D Toda lattice given by (\ref{atoda}), and also the Shabat-Yamilov lattice (see below). These formulas express integrals in terms of one unknown variable and its derivatives. The formulas are therefore valid as long as this variable is not affected by reduction or by a change of variables. Otherwise the formulas must be modified accordingly. 

Let us now demonstrate this with example of the $G_2$ Toda lattice
\begin{equation}
p_{tx}=\exp(-2 p+q), \ \ q_{tx}=\exp(3 p-2q),
\label{G2}
\end{equation}
where $w_1=p,\, w_2=q$.
Lattice (\ref{G2}) is known to be a reduction of $A_6$ Toda lattice, therefore its integrals are given by formula (\ref{scalintsm}) in which all mixed derivatives should be replaced according to system (\ref{G2}).
From formulas (\ref{transf}) and (\ref{scalintsm}) we have 
$$
\omega_i=\frac{W_{6,i}^*(\exp(p))}{W_6(\exp(p))}
$$
and thus
$$
\begin{array}{l}
\omega_6=q_x^2+3 p_{xx}-3 q_x p_x+q_{xx}+3 p_x^2, \\[1mm]
\omega_5=5\omega_{6,x},\ \
\omega_4=6\omega_{6,xx}-\omega_6^2,\ \
\omega_3=4\omega_{6,xxx}-3\omega_6\omega_{6,x}, \\[1mm]
\omega_2=\omega^*_2+\omega_{6,xxxx}-(\omega_6^2)_{xx}/2,\ \
\omega_1=\omega^*_{2,x}/2,
\end{array}
$$
where 
$$\begin{array}{l}
\omega^*_2=2 p_6+2 p_2 p_4-60 p_1^2 p_2^2+12 p_1^4 p_2-28 p_1^3 p_3-6 q_4 p_2-6 p_3 q_3-13 q_2^2 p_1^2\\
\qquad +2 p_2 q_1^4-2 p_1 (q_5-2 p_5)+14 q_3 p_1^3-10 p_2 q_2^2+4 q_1^2 q_2 p_2+26 p_1 p_2 q_3+p_3^2\\
\qquad +2 q_4 p_1^2+30 p_3 q_1 p_1^2-24 p_2^3-4 q_2 p_4+30 p_1 p_3 q_2-6 p_1 q_1^2 p_3-14 q_3 p_1^2 q_1\\
\qquad +4 q_1 q_2^2 p_1-38 p_1 q_1 q_2 p_2+18 p_1 p_4 q_1-12 q_3 p_2 q_1+2 q_1^2 p_1 q_3+36 p_2 p_3 q_1\\
\qquad -18 p_1 p_2 q_1^3-12 q_1 p_3 q_2+36 p_1 q_1 p_2^2+p_1^2 (q_1-p_1)^2 (q_1-2 p_1)^2 -2 q_2 p_1^4 \\
\qquad -2 p_2^2 q_1^2-10 q_2 q_3 p_1-14 p_4 p_1^2+34 q_2 p_2^2-16 q_1^2 q_2 p_1^2-4 p_4 q_1^2+4 q_1^3 q_2 p_1 \\
\qquad -48 p_1^3 p_2 q_1-72 p_1 p_2 p_3-4 q_1 q_4 p_1+16 q_1 q_2 p_1^3+50 p_1^2 p_2 q_1^2+68 p_2 q_2 p_1^2.
\end{array}
$$
To save space we have denoted $p_i=\p_x^i p, \, q_i=\p_x^i q$.

The other property which is common among explicitly solvable equations is presence of generalized symmetries.
The generalized symmetries of the higher d'Alembert equation can be obtained from its integrals by means of the formula
\begin{equation}
	u_\tau=\left(\frac{n-1}{2}u D_x-u_x\right)\omega+\left(\frac{n-1}{2}u D_t-u_t\right)\bar \omega,
	\label{symmetries}
\end{equation}
where $\omega$ and $\bar \omega$ are $t-$ and $x-$integrals of (\ref{2nth}).

\section{Analogs of open Toda lattices}
Previously we have raised the question of whether there are other lattices associated with equation (\ref{2nth}) which would have reasonably simple form. 
Below we present two such examples of this sort: the Shabat-Yamilov lattice and the other form of the 2D Toda lattice. We are unaware whether the connection between these lattices and equation (\ref{2nth}) has been mentioned elsewhere.

The following lattice
\begin{equation}
	w_{j,tx}=w_{j,t}w_{j,x}\left(\frac{1}{w_j-w_{j-1}}-\frac{1}{w_{j+1}-w_j}\right)
	\label{chiralchain}
\end{equation}
was introduced by Shabat and Yamilov \cite{yamil} as one the 2D anolgs of the degenerations of the Landau-Lifshitz model.
One can check that on solutions of (\ref{2nth}), the quantities
\begin{equation}
\begin{array}{l}
	w_j=\p_u \log W_j(u), \ \ j=1,\dots,n-1,
\end{array}
\label{transfc}
\end{equation}
satisfy equations of lattice (\ref{chiralchain}) along with the boundary conditions
$$
w_0=0,\ \ w_n=\infty.
$$
Lattice (\ref{chiralchain}) can therefore be viewed as an analogue of the open $A_n$ Toda lattice.
The Shabat-Yamilov lattice has some well known particular cases, for example, for $n=3$ we have the degenerate Lund-Regge (complex sine-Gordon I) system
\begin{equation}
	v_{tx}=\frac{w v_t v_x}{v w-1}, \ \ w_{tx}=\frac{v w_t w_x}{v w-1},
	\label{deglund}
\end{equation}
where $w_1=1/v,\,w_2=w$. System (\ref{deglund}) was used in \cite{sokstarts,demstarts} as a working example for demonstrating properties of  Liouville-type systems.

 Note that equation (\ref{2nth}) admits reductions that make it possible to construct 
analogues of lattice (\ref{chiralchain}) corresponding to Lie algebras $C_n, B_n,$ and possibly $D_n$. This problem will be considered elsewhere.
Instead we give one example of such lattice that is akin to the Toda $C_2$ lattice. One can verify that equation $W_5(u)=0$ admits the reduction $W_3(u)=u$. The latter equation can then be written as the system
\begin{equation}
v_{tx} = \frac{w v_x v_t }{w v-1},\ \ w_{tx}=\frac{v w_x w_t}{w v-1}-\frac{(w v-1)^3 v}{v_t^2 v_x^2},
	\label{chiraltodac2}
\end{equation}
where
$$
1/v=\p_u \log(W_1(u))=1/u,\ \ 
w=\p_u \log(W_2(u))=\frac{u_{tx}}{u_{tx} u-u_x u_t}.
$$
Yet another example of a lattice related to equation (\ref{2nth}) is given by
\begin{equation}
	v_{j,tx}=\exp(v_{j+1}-v_j)-\exp(v_j-v_{j-1}),\ \  j=1,\dots,n-1
	\label{atoda}
\end{equation}
with the boundary conditions
$$
v_0=\infty, \ \ v_n=-\infty. 
$$
This is another well known avatar of the 2D Toda lattice. The transformation relating (\ref{_todaa1}) and (\ref{atoda}) is given by
$$
v_j=w_j-w_{j-1}.
$$
On the other hand system of equations (\ref{atoda}) is related to (\ref{2nth}) via the transformation
\begin{equation*}
	v_{j+1}=-\log\left(\frac{\p}{u_{jj}} \log(W_{j+1}(u))\right),\ \ j=0,\dots,n-2,
\end{equation*}
where 
\begin{equation*}
	u_{jj}=\frac{\p^{2i}u}{\p t^j\p x^j}.
\end{equation*}
There are examples of explicitly solvable systems which seem to be related to equation (\ref{2nth}) of a particular order. Consider, for example, the equation 
\begin{equation}
	W_4(u)=0.
	\label{w4}
\end{equation}
 Introducing the variables
$$
m=-2\log(u),\ \ v=-\frac{4}{a}\, W_2(u),\ \ w=\frac{a c}{4 u} \frac{\p \log( W_3(u))}{\p u_{tx}}
$$
we may rewrite equation (\ref{w4}) as the system
\begin{equation}
m_{tx} = \frac{a}{2} v\exp(m),\ \ v_{tx} = \frac{w v_x w_t}{v w+c}, \ \ w_{tx}=\frac{v w_t w_x }{v w+c}+\frac{a}{4}(v w+c)\exp(m).
	\label{chiralsys}
\end{equation}
This system was obtained in \cite{demliouv} as a degenerate version of an S-integrable system.

The general solutions and integrals of the above systems can be easily derived from formulas (\ref{scalgensol}) and (\ref{scalints}).
Of course, the transformations given above do not exhaust all possible connections between equations of type (\ref{2nth}) and open 2D lattices.

\section*{Acknowledgments}
Author is thankful to V.E. Adler for pointing out reference \cite{yamil}, and to  V.V.Sokolov, J. H. Lee for attention to this work and many fruitful discussions.


\begin{thebibliography}{99}
\bibitem{leznov} A. N. Leznov, M. V. Saveliev, Group-Theoretical Methods for Integration of Nonlinear Dynamical Systems
[in Russian], Nauka, Moscow (1985); English transl., Birkh¨auser, Basel (1992).
\bibitem{zhibsok} A. V. Zhiber,  V. V. Sokolov, Exactly integrable
hyperbolic equations of Liouville type, { Russ. Math. Surveys} { 56}, 61--101, (2001).
\bibitem{zhib} A. M. Guryeva, A. V. Zhiber, Theoretical and Mathematical Physics, 138(3): 338–355 (2004).
\bibitem{sokstarts} V. V. Sokolov, S. Ya. Startsev, Symmetries of nonlinear hyperbolic systems of the Toda lattice type,
Theor. Math. Phys., 155(2): 802–811 (2008).
\bibitem{shabyam} A. B. Shabat and R. I. Yamilov, Exponential systems of type I and Cartan matrices, Preprint,
Bashkirian Branch, USSR Acad. Sci., Ufa (1981).
\bibitem{darb} G. Darboux, Lecons sur la theorie generale des surfaces.  / G. Darboux. Paris: Hermann, 1915. V.2.
\bibitem{yamil} A. B. Shabat, R. I. Yamilov.  To a transformation theory of two-dimensional integrable systems. Phys. Lett. A 227, N. 1-2, 15--23, (1997).
\bibitem{demstarts} D. K. Demskoi, S. Ya. Startsev, On the construction of symmetries from integrals of hyperbolic systems of equations.  Fundam. Prikl. Mat.  10,  N. 1, 29--37 (2004);  eng. transl.  J. Math. Sci. (N. Y.)  136,  N. 6, 4378--4384, (2006).
\bibitem{demliouv} D. K. Demskoi, On a class of Liouville-type systems. Theor. Math. Phys.  141,  N. 2, P. 1509--1527, (2004).
\end{thebibliography}
\end{document}